# Multiband Detection of Repeating FRB 20180916B

Ketan R. Sand[1,2] , Jakob T. Faber[1,2,3,4] , Vishal Gajjar[4] , Daniele Michilli[1,2,5,6] , Bridget C Andersen[1,2] ,
Bhal Chandra Joshi[7] , Sanjay Kudale[7] , Maura Pilia[8] , Bryan Brzycki[4] , Tomas Cassanelli[9,10] , Steve Croft[11,12] ,
Biprateep Dey[13] , Hoang John[4] , Calvin Leung[5,6] , Ryan Mckinven[9,10] , Cherry Ng[10,11,12] ,
Aaron B. Pearlman[1,2,14,21,22] , Emily Petroff[1,2,15] , Danny C. Price[4,16] , Andrew Siemion[11,12,17,18] , Kendrick Smith[19] , and
Shriharsh P. Tendulkar[7,20]
[1] Department of Physics, McGill University, 3600 rue University, Montréal, QC, H3A 2T8, Canada; ketan.sand@mail.mcgill.ca
[2] McGill Space Institute, McGill University, 3550 rue University, Montréal, QC, H3A 2A7, Canada
[3] Department of Physics and Astronomy, Oberlin College, Oberlin, OH 44074, USA
[4] Department of Astronomy, University of California, Berkeley, CA 94720, USA
[5] MIT Kavli Institute for Astrophysics and Space Research, Massachusetts Institute of Technology, 77 Massachusetts Ave, Cambridge, MA 02139, USA
[6] Department of Physics, Massachusetts Institute of Technology, 77 Massachusetts Ave, Cambridge, MA 02139, USA
[7] National Centre for Radio Astrophysics, Tata Institute of Fundamental Research, Post Bag 3, Ganeshkhind, Pune, 411007, India
[8] INAF—Osservatorio Astronomico di Cagliari—via della Scienza 5, I-09047 Selargius, Italy
[9] David A. Dunlap Department of Astronomy & Astrophysics, University of Toronto, 50 St. George Street, Toronto, ON, M5S 3H4, Canada
[10] Dunlap Institute for Astronomy & Astrophysics, University of Toronto, 50 St. George Street, Toronto, ON, M5S 3H4, Canada
[11] Berkeley SETI Research Center, University of California, Berkeley, CA 94720, USA
[12] SETI Institute, Mountain View, CA, USA
[13] Department of Physics and Astronomy and PITT PACC, University of Pittsburgh, Pittsburgh, PA 15260, USA
[14] Division of Physics, Mathematics, and Astronomy, California Institute of Technology, Pasadena, CA 91125, USA
[15] Anton Pannekoek Institute for Astronomy, University of Amsterdam, Science Park 904, 1098 XH Amsterdam, The Netherlands
[16] International Centre for Radio Astronomy Research (ICRAR), Curtin University, Bentley, WA 6102 Australia
[17] University of Manchester, Manchester, UK
[18] University of Malta, Malta
[19] Perimeter Institute for Theoretical Physics, 31 Caroline Street N, Waterloo, ON, N2S 2YL, Canada
[20] Department of Astronomy and Astrophysics, Tata Institute of Fundamental Research, Mumbai, 400005, India
Received 2021 November 2; revised 2022 April 22; accepted 2022 April 30; published 2022 June 20

## Abstract

We present a multiband study of FRB 20180916B, a repeating source with a 16.3 day periodicity. We report the detection of four, one, and seven bursts from observations spanning 3 days using the upgraded Giant Metrewave Radio Telescope (300–500 MHz), the Canadian Hydrogen Intensity Mapping Experiment (400–800 MHz) and the Green Bank Telescope (600–1000 MHz), respectively. We report the first ever detection of the source in the 800–1000 MHz range along with one of the widest instantaneous bandwidth detections (200 MHz) at lower frequencies. We identify 30 μs wide structures in one of the bursts at 800 MHz, making it the lowest frequency detection of such structures for this fast radio burst thus far. There is also a clear indication of high activity of the source at a higher frequency during earlier phases of the activity cycle. We identify a gradual decrease in the rotation measure over two years and no significant variations in the dispersion measure. We derive useful conclusions about progenitor scenarios, energy distribution, emission mechanisms, and variation of the downward drift rate of emission with frequency. Our results reinforce that multiband observations are an effective approach to study repeaters, and even one-off events, to better understand their varying activity and spectral anomalies.

Unified Astronomy Thesaurus concepts: Radio transient sources (2008); Radio bursts (1339)

## 1. Introduction

Fast radio bursts (FRBs) are millisecond-duration pulses of unknown origin (see Cordes & Chatterjee 2019; Chatterjee 2020, and Petroff et al. 2022 for a review). Their anomalously high dispersion measure (DM) suggests an extragalactic origin, but an FRB detection by the Canadian Hydrogen Intensity Mapping Experiment/Fast Radio Bursts (CHIME/FRB; CHIME/FRB Collaboration et al. 2020a) and Survey for Transient Astronomical Radio Emission 2 (STARE2; Bochenek et al. 2020) from magnetar

SGR 1935+2154 has rekindled interests in Galactic FRB sources. Recent announcements by the CHIME/FRB Collaboration et al. (2021) have led to a fivefold increase in the source count with 607 published sources detected across 110 MHz (Pastor-Marazuela et al. 2021; Pleunis et al. 2021) to 8 GHz (Gajjar et al. 2018; Michilli et al. 2018). Earlier thought to be one-off events, sources with multiple bursts, called repeaters, have been found since the discovery of FRB 20121102A (Spitler et al. 2014). Presently, 21 repeaters are known, the majority of which have been discovered by the CHIME/FRB Collaboration et al. (2019b) and Fonseca et al. (2020). Spectrotemporal studies of bursts from these sources have revealed interesting characteristic features, in particular the downward-drifting "sad trombone" effect (Hessels et al. 2019), which has been observed in the dynamic spectra of many such events.

The repeater FRB 20180916B was discovered by CHIME/FRB Collaboration et al. (2019b) and has been localized to spiral galaxy SDSS J015800.28+654253.0 at z = 0.0337

---

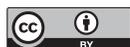

(Marcote et al. 2020). No persistent radio source has yet been discovered around the source. FRB 20180916B was found to be periodically active every 16.35 days (CHIME/FRB Collaboration et al. 2020b), with the period recently refined to be $16.33 \pm 0.12$ days with a 5.2 days activity window (Pleunis et al. 2021). It is the first FRB to be detected below 400 MHz by the Sardinia Radio Telescope (Pilia et al. 2020), the Green Bank Telescope (GBT; Chawla et al. 2020) and upgraded Giant Metrewave Radio Telescope (uGMRT; Sand et al. 2020) and is the only FRB yet to be detected below 300 MHz by the Low Frequency Array (LOFAR; Pastor-Mazuela et al. 2021; Pleunis et al. 2021) in the 110–188 MHz range. Interestingly, even after more than 100 hr of observation, spread across the entire 16.3 day period using the Deep Space Network 70 m telescope, there were no detections from the source at 2.3 GHz and 8.4 GHz (Pearlman et al. 2020). Hence, so far the Effelesberg detections at 1.7 GHz by Marcote et al. (2020) are the highest-frequency bursts reported from the source. This is dissimilar to what has been observed for FRB 20121102A, which has been detected up to 8 GHz (Gajjar et al. 2018; Michilli et al. 2018).

The source is known to have a frequency dependence in its activity, which was first observed by Aggarwal et al. (2020) in their Very Large Array detection of the source at 1680 MHz at an earlier phase beyond the activity window determined by the CHIME/FRB Collaboration et al. (2020b). This chromatic periodicity was then well established by Pastor-Marazuela et al. (2021) with their multiband observing campaign using Apertif (1370 MHz), LOFAR (150 MHz) (Pleunis et al. 2021), and CHIME/FRB (600 MHz). They not only found increased activity at earlier phases at higher frequencies, but also proposed a narrower activity window as frequency increases. Pastor-Marazuela et al. (2021), however, did observe varying burst rates from the source even during expected "peak" activity in the Apertif band and also concluded that the burst rate at LOFAR is higher than Apertif at a similar fluence threshold. This makes it imperative to undertake observations with wide bandwidth at different phases simultaneously to definitively estimate the burst-rate variation with frequency as this will put quantitative limits on the activity and energy distribution, which can help estimate some useful constraints on the emission mechanism. Notably, the repetition rate or even its status as a repeater can be biased by observing frequency. FRB 20121102A and FRB 20180916B are both good examples of this by showing no bursts at low (i.e., less than 400 MHz) and high frequencies (i.e., more than 1.7 GHz), respectively. Hence the simultaneous wideband study of repeaters or FRBs, in general, is the most efficient strategy to understand their enigmatic properties across large frequencies to eventually reveal possible progenitor scenarios.

The band-limited nature of repeaters makes it essential to have high-resolution simultaneous observations with a large bandwidth that can help detect variations in the spectro-temporal and even polarimetric structures of the bursts at different frequencies. The study of these contemporaneous frequency-dependent variations, as well as long-term changes with time, are paramount in investigating the origins of FRBs. FRB 20121102A was one of the first repeaters to be coherently corrected for dispersion, allowing unprecedented high-time-resolution studies, revealing drifting bursts and an unusually high rotation measure (RM; Gajjar et al. 2018; Michilli et al. 2018; Hessels et al. 2019). Subsequent observations have shown a considerable decrease in the RM value with time (Hilmarsson et al. 2021). This variation is similar to what has been observed for Galactic center magnetar PSR J1745-2900 (Desvignes et al. 2018). On the other hand, FRB 20180916B has a nominal RM and has shown a gradual variation in RM value over the two year timescale (Pleunis et al. 2021). These variations may be due to changes in the viewing geometry or the result of drastic variations in the circumburst environment facilitating magnetoionic disparities. Interestingly, both of these FRBs show a flat polarization position angle (PPA) that can be indicative of emissions at higher altitudes due to relativistic shock (Beloborodov 2017; Metzger et al. 2019). Contrarily, observation of microstructures challenges this idea, as in the case of FRB 20180916B, where such structures have been seen down to 2–3 $\mu$s (Nimmo et al. 2020) at 1700 MHz, hinting at emission much closer to the magnetosphere. Studying the evolution of such structures with frequency will aid in constraining the emission region and help to understand the local environment. Another aspect of repeater burst morphology that remains a mystery is the "sad trombone" effect, which describes a linear decrease in burst features with frequency and has been observed for FRB 20121102A and other repeaters; it will be interesting to undertake a comparative study of this behavior between frequencies.

Here, we present results from the observation campaign of FRB 20180916B using the uGMRT, GBT, and CHIME/FRB. This paper includes a detailed analysis of uGMRT detections of FRB 20180916B (Sand et al. 2020) along with bursts during simultaneous GBT observations in the 600–1000 MHz range; we report the first detections of the source in the 800–1000 MHz range. We also report one CHIME/FRB detection within a 0.04 activity phase of these detections.[23] In Section 2 we present details about the observations and instruments used. In Section 3 we provide the methodology that was used in the analysis. In Section 4 we discuss the implications of our results. Lastly, we provide concluding remarks in Section 5.

## 2. Observations

The source was observed using three telescope facilities, uGMRT, GBT, and CHIME/FRB. These observations were undertaken on 2020 March 23 and 24 (UTC) (see Table 1). One of the sessions for uGMRT was simultaneous with the GBT session (see Figure 1). The CHIME sessions selected are the transits of the source on the two days corresponding to our observations.

### 2.1. Giant Metrewave Radio Telescope

The uGMRT provides excellent capabilities for detecting FRBs at lower frequencies (<500 MHz) across a large instantaneous bandwidth of 200 MHz (Sand et al. 2020; Marthi et al. 2020; Pleunis et al. 2021). We observed FRB 20180916B during the activity window near the peak of its activity in the CHIME/FRB frequency range (400–800 MHz; CHIME/FRB Collaboration et al. 2020b) through a Target-of-Opportunity Director's Discretionary Time (DDT) proposal (Project code: DDTC 127). These observations were carried out at the Band 3 (250–500 MHz) of the uGMRT on 2020 March 23 and 2020 March 24 UTC (see Table 1). We formed tied-array beams using 19 antennas, which included 13 antennas from the central

---

[23] This CHIME detection was previously published in Pleunis et al. (2021).

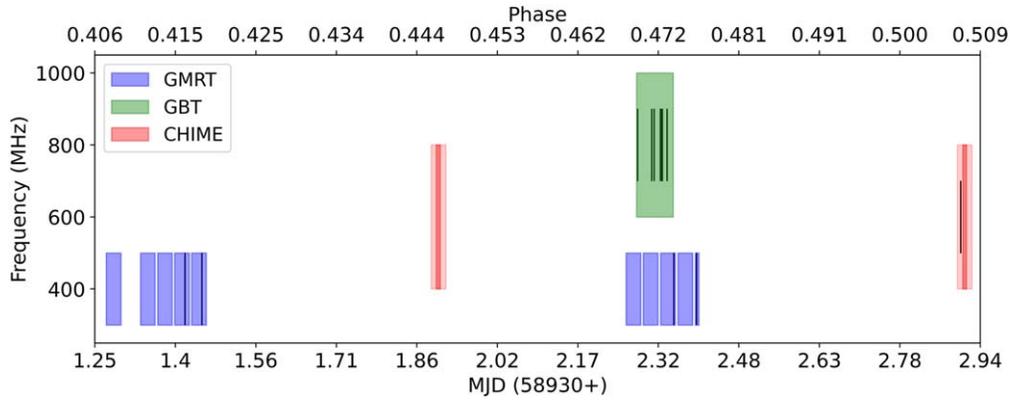

**Figure 1.** Observation epochs (in MJD and phase) and frequency coverage for all telescopes. Bursts detected are shown with black solid lines for our semisimultaneous multiband observations. The lighter shade in CHIME/FRB observations represents transit time in sidelobes (end-to-end 40 minutes) and the darker region shows transit time through the beam FWHM (~10 minutes). Phase has been calculated by folding at the period of 16.33 days with reference to MJD 58369.40 (Pleunis et al. 2021).

**Table 1**
Table of Observations and Detections

| Telescope | Frequency Range (MHz) | Start Time in UTC | Duration of Observation (Hrs) | Bursts |
|---|---|---|---|---|
| GMRT | 300–500 | 2020-3-23 06:31:19.566 | 0.67 | |
| | | 2020-3-23 08:5:00.605 | 0.67 | |
| | | 2020-3-23 08:05:00.605 | 0.67 | |
| | | 2020-3-23 08:52:04.546 | 0.67 | |
| | | 2020-3-23 09:38:44.327 | 0.67 | GMRT A |
| | | 2020-3-23 10:25:18.740 | 0.67 | GMRT B |
| | | 2020-3-24 06:17:57.000 | 0.67 | |
| | | 2020-3-24 07:04:55.732 | 0.67 | |
| | | 2020-3-24 07:52:23.832 | 0.67 | GMRT C |
| | | 2020-3-24 08:40:05.354 | 0.67 | |
| | | 2020-3-24 09:27:17.348 | 0.17 | GMRT D |
| CHIME/FRB | 400–800 | 2020-3-23 21:22:40 | 0.67 | |
| | | 2020-3-24 21:26:40 | 0.67 | CHIME A |
| GBT | 600–1000 | 2020-3-24 06:46:26.440 | 1.67 | GBT A-G |

**Note.** Columns list telescope, frequency range for each session, start time of a session (in UTC), duration of a given session in hours, and detected bursts for each session, respectively. The Transit Time for CHIME/FRB also includes sidelobes, as shown in in Figure 1.

square and the first two antennas of each arm of the array. We used 3C48 for initial phasing and power equalization for all antennas. The uGMRT wideband back end is capable of recording four simultaneous beams (Reddy et al. 2017). We utilized one of the beams in voltage mode and coherently dedispersed incoming baseband voltages to a DM of 350 pc cm$^{-3}$. We recorded 2048 dispersion-corrected subbands with a temporal resolution of 81 $\mu$s. We also recorded another beam in a standard phased-array mode as a backup, which was not utilized in these analyses. We took 40 minute scans of the source followed by observation of 3C48 in between for phase adjustments; this can be seen as gaps between scans in Figure 1. The high-time-resolution filterbank data products for each scan were rid of any bright narrowband and broadband (zero DM) interference using `gptool` (Susobhanan et al. 2021) before converting them to `SIGPROC` (Lorimer 2011) formatted standard filterbank files.

The treated filterbank files were then searched for single-pulse candidates using `SPANDAK`,[24] which uses a GPU-based tool named `HEIMDALL`[25] (Barsdell et al. 2012) as the main kernel. This pipeline is similar to the one used by Gajjar et al. (2018) and Pilia et al. (2020) for detecting FRBs at 8 GHz and 300 MHz, respectively. To flag any remaining RFI, the `gptool`-processed filterbank files were further checked for "bad" channels using `PRESTO`[26] `RFIFIND` algorithm. Then the data were searched using `HEIMDALL` for dedispersed pulses across a DM range of 0–500 pc cm$^{-3}$. Candidates generated were then scrutinized by a Python-based classification algorithm which crudely filtered out potential astrophysical candidates from noise spikes and RFI, taking into consideration maximum candidates per second (set to four in our case), signal-to-noise (S/N) thresholds (6$\sigma$ for GMRT), S/N-versus-DM distribution, and S/N-versus-boxcar width. We searched boxcar widths from 1 to 2048 time bins, where each time bin had a temporal resolution of 81 $\mu$s. The pipeline then extracted these shortlisted candidates from the original filterbank file, dedispersed them around their reported DMs, and plotted dynamic spectra and burst profile across its time duration. The plots generated were then visually inspected. More details on the pipeline will be presented elsewhere. We identified four bursts, labeled as GMRT A–D (see Figure 3), in the GMRT data after looking at more than 20,000 candidates generated by the pipeline. These bursts were then extracted as archive files at the native resolution of the filterbank files (frequency resolution ∼98 kHz, time resolution ∼81.92 $\mu$s) using DSPSR for further analysis. One of the 40 minute scans from 2020 March 23 was corrupted due to interference and hence was not included in the further analysis (see Figure 1).

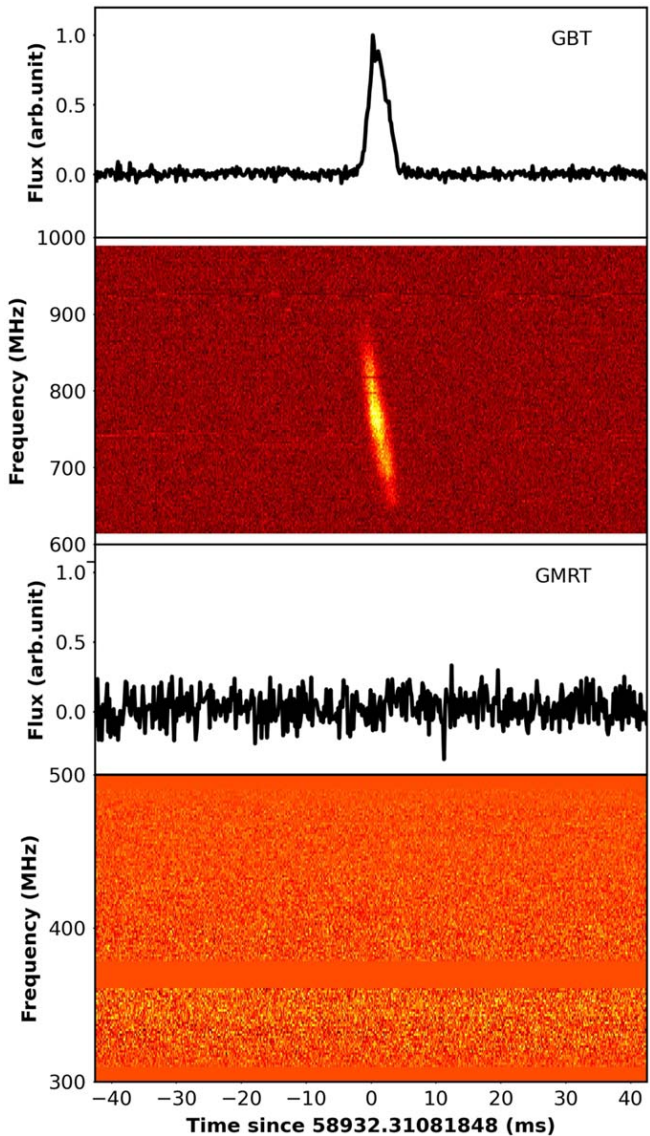

**Figure 2.** Simultaneous GBT and GMRT observations of a burst (GBT B) only detected at the GBT. The top panels in both plots show integrated and dedispersed time series while the second panels in both plots show dedispersed dynamic spectra at a DM of 348.8 pc cm$^{-3}$. We clearly see the band-limited nature of these bursts and confirm that the burst energy is within our observed frequency range.

### 2.2. Green Bank Telescope

The observations with the GBT were conducted on 2020 March 24 (Table 1) across 680–920 MHz using the Prime Focus 1 receiver, and data were collected with the Breakthrough Listen digital back end (BLDB; MacMahon et al. 2018). The BLDB is a highly flexible, state-of-the-art, GPU-enabled 64-node compute cluster at the GBT. The BLDB is primarily deployed at the GBT to conduct one of the most comprehensive searches for intelligent life in the universe (Isaacson et al. 2017; Worden et al. 2017; Gajjar et al. 2019). BLDB records 8-bit complex voltages with up to 12 GHz of instantaneous bandwidth across 64 compute nodes, with each node recording 187.5 MHz of bandwidth. We configured it to record across 614–989 MHz (375 MHz of total bandwidth), covering the 680–920 MHz receiver bandwidth, utilizing two compute nodes. These observations were done in simultaneity with one of the uGMRT epochs (see Figure 1). We recorded complex voltages into the Green Bank ultimate pulsar processing instrument (GUPPI) format with the native sampling time of 341 ns with 128 polyphase coarse channels (see Lebofsky et al. 2019). These raw voltages were converted to total-intensity SIGPROC filterbank files with a spectral and temporal resolution of ∼350 kHz and 349 $\mu$s, respectively, for searching. A similar search for FRBs was done here as described for GMRT bursts using SPANDAK, with a few exceptions. First, we did not treat the files for RFI before converting them to SIGPROC filterbank files. Second, we performed a subbanded search (dividing the band into eight subbands) to identify low-intensity and narrowband (also referred to as "smudgy" bursts. We searched boxcar widths from 1 to 2048 time bins, where each time bin had a temporal resolution of 350 $\mu$s. This search (above 6$\sigma$) revealed seven real candidates (GBT A–G; Figure 3) after an initial visual examination of around 5000 candidates, followed by a cross-referencing of candidates detected in individual subbands to pick out coincident signals that extended beyond a single subband. Once we had identified all bursts in the data, we then extracted their raw voltages and coherently dedispersed them around a fiducial DM of 349 pc cm$^{-3}$ using DSPSR to produce archive files of desired time and frequency resolutions depending on the analysis involved. After a detection of a burst at the GBT, we carefully again reexamined uGMRT data around the same epoch (correcting for dispersion delay). Figure 2 shows a clear detection of a burst at the GBT while, at the same instant, uGMRT does not show any indication of any dispersed burst. This confirms that, similar to FRB 20121102A (Law et al. 2017), bursts from FRB 20180916B are highly spectrally limited.

### 2.3. CHIME/FRB

CHIME is a radio interferometer observing in the 400–800 MHz range. The CHIME/FRB back end (CHIME/FRB Collaboration et al. 2018) provides the capability of detecting and studying the FRB population (CHIME/FRB Collaboration et al. 2021). With its real-time pipeline, CHIME/FRB continuously monitors the large (∼200 square degrees) field of view of the telescope for impulsive, dispersed signals. RFI is automatically discarded with a dedicated machine-learning algorithm, and a few seconds of total-intensity data are stored around signals of interest for further analysis (CHIME/FRB Collaboration et al. 2021). These data have a time resolution of 0.98 ms and a frequency resolution of 24.4 kHz.

One burst (Figure 3, bottom right) from FRB 20180916B was detected by the CHIME/FRB within one day of the detections from the GMRT and GBT summarized in Table 2. We calculated its peak flux and fluence by using steady sources as calibrators following previous work presented by the CHIME/FRB Collaboration et al. (2019a) and the CHIME/FRB Collaboration et al. (2021).

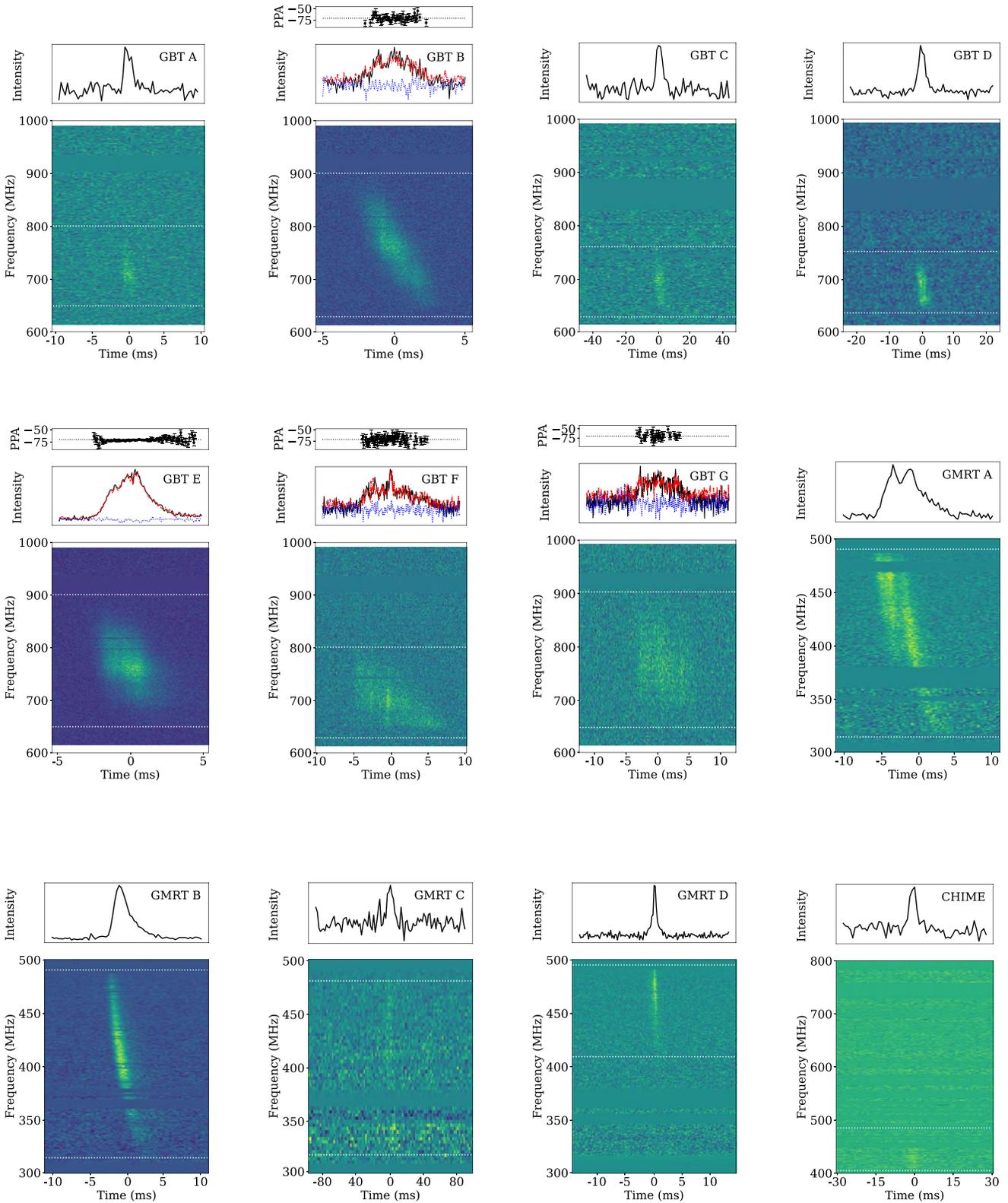

**Figure 3.** Waterfall plots of all 12 reported bursts. The top panel shows the integrated profile with bottom panel showing flux as a function of time and frequency. Each burst has been dedispersed to a DM of 348.8 pc cm$^{-3}$, the structure-maximizing DM for the highest S/N burst in our data set (GBT E; see Section 3.1 ), except bursts GBT D (349.8 pc cm$^{-3}$), GMRT C (350 pc cm$^{-3}$), and CHIME (349.5 pc cm$^{-3}$), which were too weak to be seen around this DM. For GBT bursts B, E, F, and G, circular (blue) and linear (red) polarization and PPA referenced to infinite frequency are plotted against the respective time series. The white dotted lines show the extent of the frequency range summed to get the integrated profile.

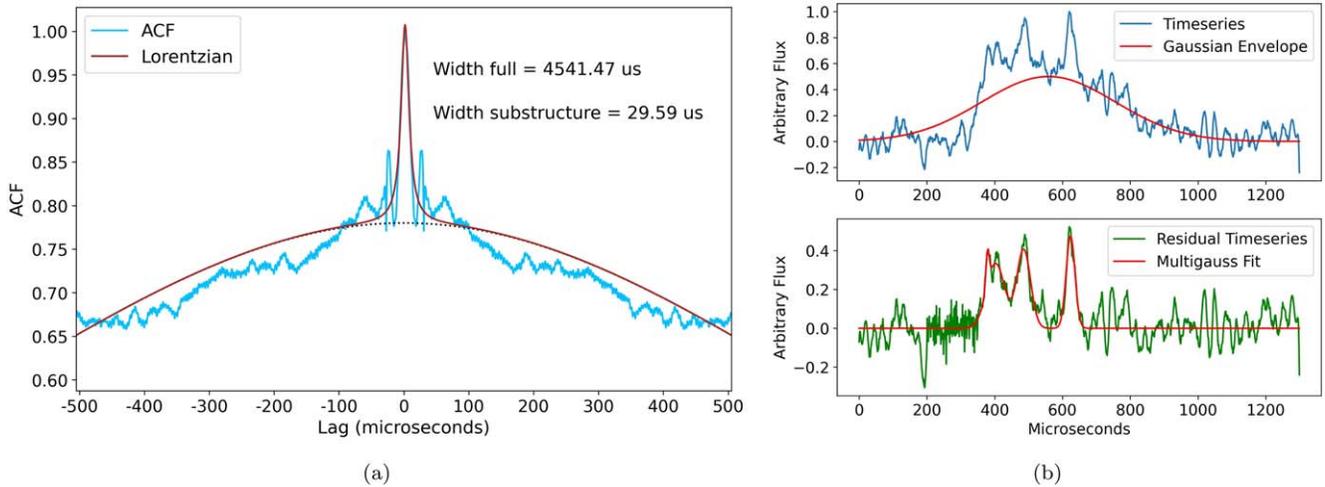

(a)                                                                                                          (b)

**Figure 4.** Microstructure analysis for GBT F burst. (a) Time-series autocorrelation of GBT F to study microstructure. The width of the structure is estimated to be 30 $\mu$s using a Lorentzian fit. (b) The 4 $\mu$s time series of the same burst with envelope-subtracted residual showing clear structures.

**Table 2**
Burst Properties of Detected Bursts from FRB 20180916B

| Burst | MJD (58932+) | Peak S/N Full/$\Delta\nu$ | $\Delta\nu$ (MHz) | Width[a] (ms) | Peak Flux Full/$\Delta\nu$ (Jy) | Fluence Full/$\Delta\nu$ (Jy ms) | $t_{peak}$ (ms) |
|---|---|---|---|---|---|---|---|
| GBT A | 0.2827216 | 7.85/18.12 | 46 | 2.8(1.5) | 0.29(2)/1.51(9) | 0.35(21)/1.8(1.1) | 0.33 |
| GBT B | 0.3093268 | 37.17/46.05 | 179 | 3.13(51) | 2.04(25)/3.19(39) | 2.72(77)/4.2(1.2) | 0.16 |
| GBT C | 0.3141119 | 6.16/11.74 | 81 | 7.0(1.6) | 0.12(1)/0.39(1) | 0.35(8)/1.19(27) | 1.3 |
| GBT D | 0.3260644 | 8.51/18.15 | 86 | 2.4(1.5) | 0.23(1)/0.95(3) | 0.24(15)/0.98(62) | 0.65 |
| GBT E | 0.3291835 | 80.22/97.75 | 146 | 3.28(61) | 4.39(55)/7.03(87) | 6.1(1.9)/9.8(3.1) | 0.16 |
| GBT F | 0.3297834 | 14.33/24.84 | 118 | 10.3(2.6) | 0.54(3)/1.47(9) | 2.41(75)/6.5(2.1) | 0.33 |
| GBT G | 0.3388846 | 13.69/17.99 | 177 | 8.2(1.9) | 0.52(3)/0.95(5) | 1.83(53)/3.33(97) | 0.33 |
| GMRT C | 0.3516781 | 9.85/9.85 | <200 | <8.4(2.3) | 0.43(1)/0.43(1) | 1.55(45)/1.55(45) | 1.9 |
| GMRT D | 0.3952298 | 20.48/46.56 | 74 | 2.7(1.0) | 1.29(3)/4.98(12) | 1.49(59)/5.7(2.3) | 0.8 |
| CHIME[b] | 0.8983427 | 13.0 | <50 | 3.1(3) | >0.4(2) | >1.6(6) | 0.98 |
| | MJD (58931+) | | | | | | |
| GMRT A | 0.4200371 | 24.58/25.65 | 171 | 5.82(73) | 2.53(15)/2.85(17) | 6.2(1.1)/7.1(1.2) | 0.33 |
| GMRT B | 0.4523516 | 67.39/67.39 | >200 | 2.46(47) | 6.92(42)/6.92(42) | 7.2(1.9)/7.2(1.9) | 0.33 |

**Notes.** Columns list arrival MJDs (barycentered and referenced to infinite frequency), peak S/N, fractional bandwidth of burst emission, measured burst width, peak flux, fluence, and temporal resolution of underlying data products used for these measurements, respectively.
[a] All burst widths were estimated using burst profiles integrated over the full observing band.
[b] Properties estimated using CHIME/FRB pipelines as described in Pleunis et al. (2021).

## 3. Analysis

The analysis presented here is used in the characterization of GMRT and GBT bursts. Details regarding the estimation of properties for the CHIME/FRB burst have been presented in Pleunis et al. (2021).

### 3.1. Burst Properties, Scattering, and Microstructures

Both the GMRT and GBT bursts exhibit complex morphology. As seen in Figure 3, GMRT A shows a downward-drifting substructure, with GMRT B displaying what seems to be a scattering tail but can also be some unresolved structures. GBT B and GBT E both show drift in the spectra, with GBT F exhibiting microstructures (Figure 4). These structures will be hindered if we dedisperse as per S/N; hence, to estimate the structure-maximizing DM we used the `DM phase`

(Seymour et al. 2019) package.[27] This algorithm maximizes the coherent power of the bursts. To estimate the best DM each burst was dedispersed around a trial DM (348.7 pc cm$^{-3}$); the `DM phase` package then tested from −6 to +6 values around the trial DM with steps of 0.01 to ascertain the structure-maximizing DM. We got DM values for all bursts within 348.6–350.2, which is consistent with the DM reported for this FRB. For GBT E, the highest S/N burst in our data set, we found a structure-maximizing DM of 348.9 ± 0.1. We then dedispersed all other bursts around this DM range (348.8–349.0 pc cm$^{-3}$) and, after a visual examination, found that a DM of 348.8 pc cm$^{-3}$ seems to maximize structures for all the bursts in our data set (see Figure 3). Thus, this DM was then used in all the analysis reported herein.

---

[27] https://github.com/danielemichilli/DM_phase

**Table 3**
Spectrotemporal and Polarization Properties for Selected Bursts Detected from the GBT and GMRT

| Burst | $RM_{obs}$ (rad m$^{-2}$) | $PPA_\infty^{mean}$ (deg) | $\Delta PPA_\infty$ (deg) | $\tau_s$ (ms) | $\Delta\nu_{DISS}$ (kHz) | Drift Rate (MHz/ms) |
|---|---|---|---|---|---|---|
| GBT B | −117.5(5) | −71.4(8) | 6 | ... | 2.6(6) | −35.7(8) |
| GBT E | −117.7(2) | −71.30(2) | 4 | <1.16(31) | 4.2(2) | −18.6(8) |
| GBT F | −117.4(9) | −71.4(4) | 7 | ... | ... | −8(1) |
| GBT G | −117.2(9) | −69(1) | 8 | ... | ... | −10(2) |
| GMRT A | ... | ... | ... | 2.51(37) | ... | −13.6(6) |
| GMRT B | ... | ... | ... | 1.67(47) | ... | ... |

**Note.** Columns, from left to right, report burst name, measured RM, mean PPA at infinite frequency in barycentric reference frame, standard deviation on the measured PPA, scattering timescale, measured Diffractive interstellar scintillation (DISS) bandwidth, and drift rate of downward-drifting emission. We report the upper limit on the scattering timescale of GBT E. The RM reported here is corrected for an ionospheric contribution of +0.1 rad m$^{-2}$ (see Section 3.3).

To estimate width and scattering, we used `bilby`[28] a Bayesian inference library, to perform a nested sampling fit of multiple exponentially modified Gaussians, one for each subburst over the burst profile. These profiles were generated by frequency-averaging over the entire band. The counting of the number of subbursts for every burst was performed based on the appearance of pronounced peaks and valleys in each burst profile. We performed a nested sampling fit of the corresponding multiple exponentially modified Gaussians to each burst. For around half of our bursts (namely, GBT A, GBT C, GBT D, GMRT C, and GMRT D), our fit did not convincingly converge for exponentially modified Gaussians while fitting the burst profile. Hence, for these bursts we only used multiple Gaussians to fit burst profiles. For the remaining bursts (namely, GBT B, GBT E, GBT F, GBT G, GMRT A, and GMRT B), we fitted multiple exponentially modified Gaussians. We defined the width of the burst as the separation between half-maximums of the leading edge of the first subburst and the trailing edge of the last subburst from the corresponding fitted model. Measured widths are listed in Table 2. It should be noted that, although for a few of the bursts we were able to obtain a good fit for an exponential tail, not all of them are likely to arise due to scattering. Some of them are likely due to unresolved subbursts. Hence, we only report the possible scattering timescale for three of our high-S/N bursts with best exponential fit in Table 3. Our scattering timescale measurements for GMRT bursts, particularly GMRT B with ∼1.7 ms, which shows a clear exponential tail, is in agreement with what has been observed for the source around similar frequencies by Chawla et al. (2020).

FRB 20180916B has already been reported to exhibit submicrosecond structure at 1.7 GHz (Nimmo et al. 2020). We also performed autocorrelation function (ACF) analysis to investigate microsecond structure in our bursts detected from the GBT at 800 MHz, where we had retained baseband voltages. Unfortunately, such a study was not possible to perform for the bursts detected from the GMRT due to insufficient S/N and poor temporal resolution of our data products. In order to look for microstructures, we again coherently dedisperse baseband voltages and produce higher temporal resolution data products of 4 μs. We performed an ACF on the dedispersed and frequency-averaged time series for one of the bursts, GBT F, which was already showing noticeable substructures (see Figure 3). As shown in Figure 4(a), the ACF showed a significant peak. The measured widths of the ACF from a Lorentzian fit is around 30 μs. Such microstructures are readily seen in Figures 4(b), once a Gaussian envelope of the overall burst shape is removed.

To estimate the peak flux for our detected bursts, we used the radiometer equation (Lorimer & Kramer 2004), which can be expressed as

$$S_{peak} = \beta \frac{(S/N)_{peak} T_{sys}}{G \sqrt{n_p \Delta\nu t_{peak}}}. \quad (1)$$

Here, $T_{sys}$ is the system temperature (receiver + sky), $\Delta\nu$ is the bandwidth, $G$ is the gain of the telescope, $\beta$ is the digitization factor (quantum efficiency), $n_p$ is the number of polarization channels, and $t_{peak}$ is the temporal resolution of the burst profile. The $(S/N)_{peak}$ was calculated after normalizing time series by subtracting the mean of an off-pulse region from an on-pulse region, where the on-pulse region was defined by visually inspecting the burst profile for each burst. For the GBT,[29] $\beta \sim 1$, $T_{sys} = 38$ K (29 K + 9 K), $G = 2$ K Jy$^{-1}$, $\Delta\nu = 375$ MHz, and $n_p = 2$. In case of the GMRT, an additional antenna term $\sqrt{\delta N_{ant}(N_{ant} - 1)}$ was multiplied in the denominator, where $N_{ant} = 19$ corresponds to the number of antennas employed to create a phased array and $\delta$ accounts for the linear degradation of phasing that will lead to loss of sensitivity. For example, by carefully monitoring the phasing at the beginning and at the end of a single 40 minute long scan, we noticed a decrease in $\delta$ from 0.8 to 0.65, which we assumed to follow a linear trend to affect flux estimates for different bursts detected across the length of a scan. Other values[30] were taken to be $T_{sys} = 218$ K (165 K + 53 K), $n_p = 2$, $G = 0.38$ K Jy$^{-1}$, and $\Delta\nu = 200$ MHz. The sky temperature in case of the GMRT (53 K) and GBT (9 K) was estimated using the 408 MHz sky temperature map from Haslam et al. (1982) and using a spectral index of −2.55 (Remazeilles et al. 2015), extending to central frequencies of 400 MHz and 800 MHz, respectively. As seen in Figure 3, many bursts exhibit spectrally limited emission, thus we measured peak fluxes and fluences from full bandwidth and fractional bandwidth for each burst. This fractional bandwidth was determined by a Gaussian fit to each burst spectrum (Zhang et al. 2018). The fluences for full and fractional bandwidths were calculated by summing over FWHM temporal width and averaging over full and fractional bandwidth, respectively. Table 2 summarizes these properties for reported bursts from the GMRT, GBT, and CHIME.

---

[28] https://lscsoft.docs.ligo.org/bilby/index.html

[29] https://science.nrao.edu/facilities/gbt/proposing/GBTpg.pdf

[30] http://indrayani.ncra.tifr.res.in/~secr-ops/sch/c41webfiles/gtac_41_status_doc.pdf

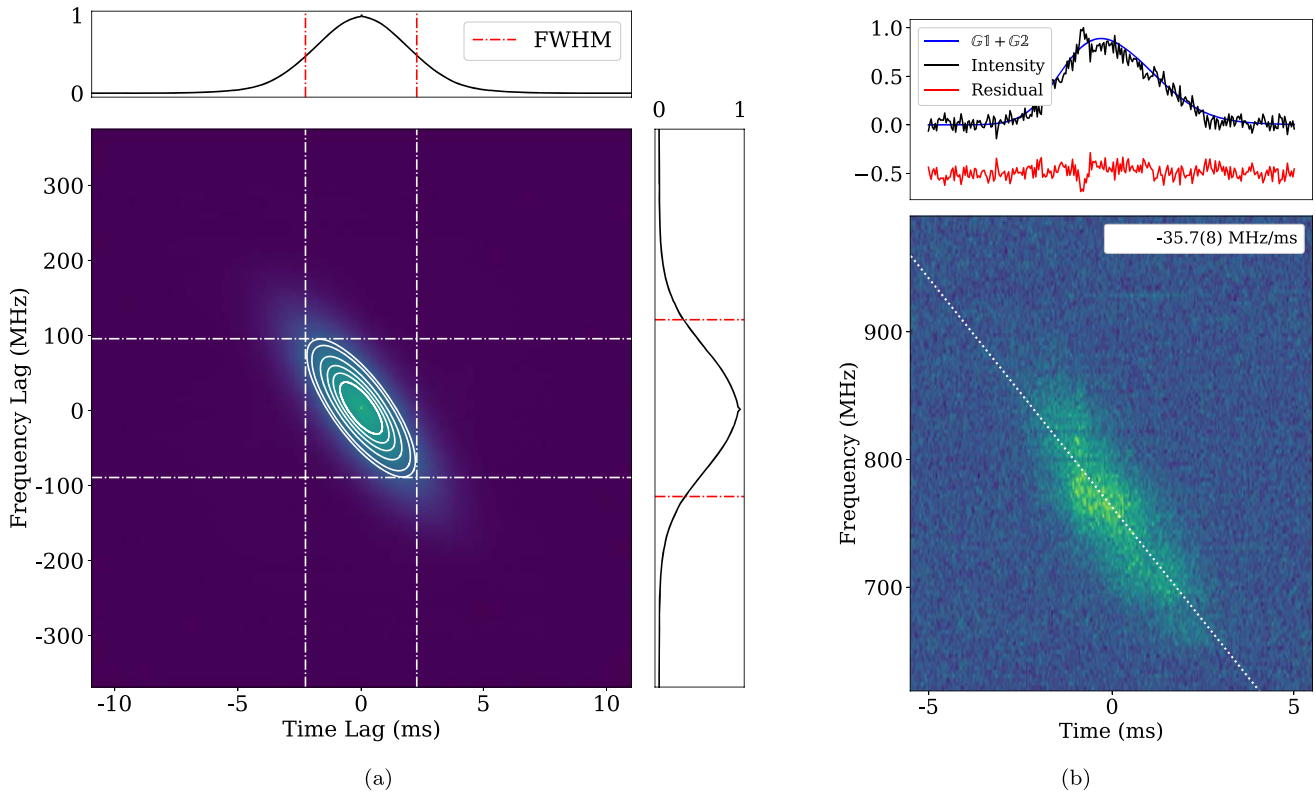

**Figure 5.** Drift-rate estimation of burst GBT B from FRB 20180916B. (a) 2D ACF of GBT B with overlaid fitted ellipses using `photutils`. (b) Dynamic spectrum with overplotted dashed line revealing the drift rate, as well as the burst profile with fitted double exponentially modified Gaussians.

### 3.2. Drift-rate Estimation

One of the most notable morphological features present in repeating FRBs is subpulse drifting (also known as the "sad trombone" effect). Drifting appears as the downward progression of the observed frequency of the subpulses comprising a burst as a function of time. This can appear in bursts that are cohesive in time (exhibiting no features between which the power drops to the noise level), i.e., the majority of bursts in this data set, as well as in bursts that are made up of discrete subbursts, such as GMRT A. One method that has been used thus far to calculate drift rates in FRBs involves transforming the dynamic spectrum to spectrotemporal lag space by taking its 2D ACF, followed by using Monte Carlo sampling to fit an ellipse to the delay spectrum, and finally deriving the drift rate from the tilt of its semimajor axis (Gourdji et al. 2019; Hessels et al. 2019). Here we present the results achieved by using a slightly different method that, to the best of our knowledge, has not been used for measuring drift rates in the past.

The method implemented here is achieved through elliptical isophote analysis using a package within `Astropy` called `photutils`, which, similar to the standard method, is performed on the 2D autocorrelation of the dynamic spectrum. The isophotometry algorithm iteratively samples pixels along trial ellipses of varying ellipticity and position angle in the ACF for a given semimajor axis length, hoping to identify pixels with similar intensity which indicate a good description of (or fit to) an isophote. We have chosen to avoid sampling for small semimajor axis values due to significantly higher intensities in the bins that occupy the center of the ACF. As seen in Figure 5(a), the fitted elliptical isophotes were plotted to the

FWHM of the ACF for burst GBT B. Figure 5(b) shows the original waterfall plot, over which a dotted white line has been plotted with the averaged slope of all fitted isophotes in the ACF. The drifting estimates vary for GMRT and GBT bursts (Table 3). GMRT A has a clear subburst drift of around $-13.6$ MHz ms$^{-1}$, which is less than the drift observed for brighter GBT bursts, in particular GBT B with a drift of around $-36$ MHz ms$^{-1}$. This linear variation with frequency (see Figure 7(a)) is similar to what has been observed for repeaters, notably FRB 20121102A, which showed a higher drift rate in its 8 GHz detections (Gajjar et al. 2018) compared to its $L$-band detections (Hessels et al. 2019).

### 3.3. Polarimetry

We carried out measurements of the polarization properties of the GBT bursts as we only recorded total intensity for the GMRT bursts. To measure polarization, we first extracted the baseband raw voltage data for four GBT bursts (GBT B, GBT E, GBT F, and GBT G) with sufficiently high S/N values and coherently dedispered it at the previously measured DM of 348.6 pc cm$^{-3}$. These products were stored in full-Stokes `PSRFITS` format at resolutions of 40 $\mu$s and 97.66 kHz for all four bursts.

The remaining bursts detected from the GBT did not offer sufficient S/N to reliably extract any polarization information. During the observations, a noise diode scan was performed at the beginning of the observing run. This noise diode scan on the source was used for polarization calibration. We used the `PSRCHIVE`'s `pac` routine to calibrate our `PSRFITS` file for each burst.

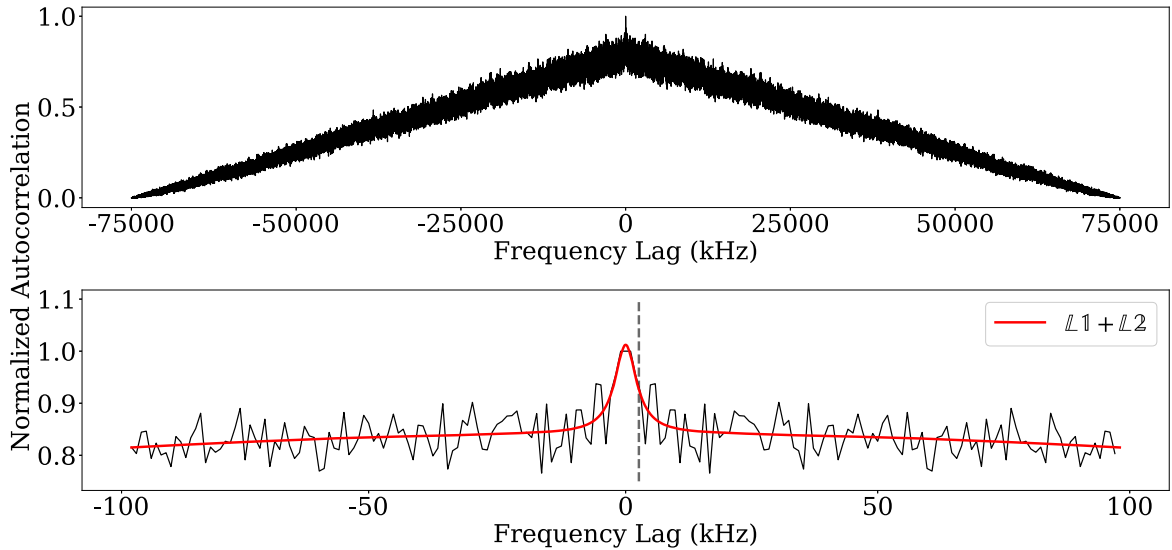

**Figure 6.** Scintillation bandwidth measurement for GBT B burst from FRB 20180916B. Top: ACF of 75 MHz of bandwidth, centered around 794 MHz (the centroid frequency). Two Lorentzian functions were fit to the central peak and broader underlying curve in the ACF. Bottom: the Lorentzian fit to the central peak, with FWHM shown with a dotted vertical line at 2.6(6) kHz, which corresponds to the scintillation (decorrelation) bandwidth. This scintillation bandwidth is in agreement with the predicted bandwidth of 2.68 kHz from NE2001 (Cordes & Lazio 2002).

To measure the RM of each burst, we utilized two algorithms in tandem, both of which are implemented by the rmfit[31] tool within PSRCHIVE. The first is a brute-force calculation of fractional linear polarization for a series of trial RM values. The value achieved with this method is used as an initial guess for the second algorithm, which performs an iterative refinement of the differential PPA. This involves dividing the band in two and calculating the weighted differential polarization angle between each half. For the latter method, as long as the weighted differential PPA is greater than its uncertainty, the data are iteratively corrected for the corresponding RM trial value until the weighted differential PPA is less than its uncertainty. We also measured the RM contribution from ionosphere using ionFR[32] (Sotomayor-Beltran et al. 2013), which was found to be +0.1 rad m$^{-2}$ around the arrival time of all the GBT bursts. Table 3 reports the RM values after correcting for this ionospheric contribution. The RM values we obtain are slightly less than the values first measured for this source (CHIME/FRB Collaboration et al. 2019b), but remain consistent with the most recent measurements (Nimmo et al. 2020) within an uncertainty of 1$\sigma$.

The effect of Faraday rotation was corrected to the measured RM values of each individual burst, all of which are recorded in Table 3, with an average value of −117.6 rad m$^{-2}$. The correction reveals nearly 100% linear polarization and stable PPA curves across pulse phase, as shown in Figure 3. We also noticed flat PPA during the saddle emission between subbursts, which confirms findings recently reported by Nimmo et al. (2020). The linear polarization plotted over bursts GBT B, E, F, and G adheres to the complex substructures of each burst—confirming that these substructures are indeed physical.

### 3.4. Scintillation Studies

We investigated the scintillation (decorrelation) bandwidth for GBT B and E. We required a high frequency resolution to resolve the bandwidth predicted by the NE2001 Galactic electron density model (Cordes & Lazio 2002), approximately 2.94 kHz for an observing frequency of 800 MHz. For the GMRT bursts our recorded frequency resolution of 98 kHz was insufficient to observe any scintillation effect. To perform the analysis for GBT bursts, archive files were re-generated from the baseband data with frequency and time resolutions of 0.98 kHz and 1.00 ms, respectively. Using the on-pulse spectrum, we identified the centroid of each burst (the point of peak brightness) by fitting a Gaussian curve to its spectrum. We then excised 75 MHz of bandwidth centered on the centroid frequency (roughly 794 MHz for GBT B and 782 MHz for GBT E) from the dynamic spectrum, which we used to calculate an ACF. The bandwidth was cropped around the brightest part of the burst to improve the S/N in the ACF. Finally, as shown in Figure 6, a stacked double Lorentzian was fit to the ACF—the first to the underlying curve of the ACF, and the second to the central peak. The FWHM of the Lorentzian fit to the central peak reveals the scintillation bandwidth. For GBT B, we calculate a scintillation bandwidth of 2.6(6) kHz for a centroid frequency of 794 MHz, whereas for GBT E, despite the burst being centered slightly lower in frequency, we measured a bandwidth of 4.2(2) kHz. The scattering estimate obtained from the average bandwidth between GBT B and GBT E is around 0.04 ms. This differs quite drastically from the scattering timescale measured from the burst profile of GBT E, which leads us to believe that the exponential fall-off in intensity visible in the profile is a morphological feature rather than the product of interstellar scattering.

### 4. Discussion

#### 4.1. Spectrotemporal Properties

Repeaters are known to have complex morphologies. FRB 20121102A is an excellent example, where such multi-component downward-drifting structures have been observed as high as 8 GHz (Gajjar et al. 2018; Hessels et al. 2019). FRB 20180916B also exhibits this property and we too observe

---

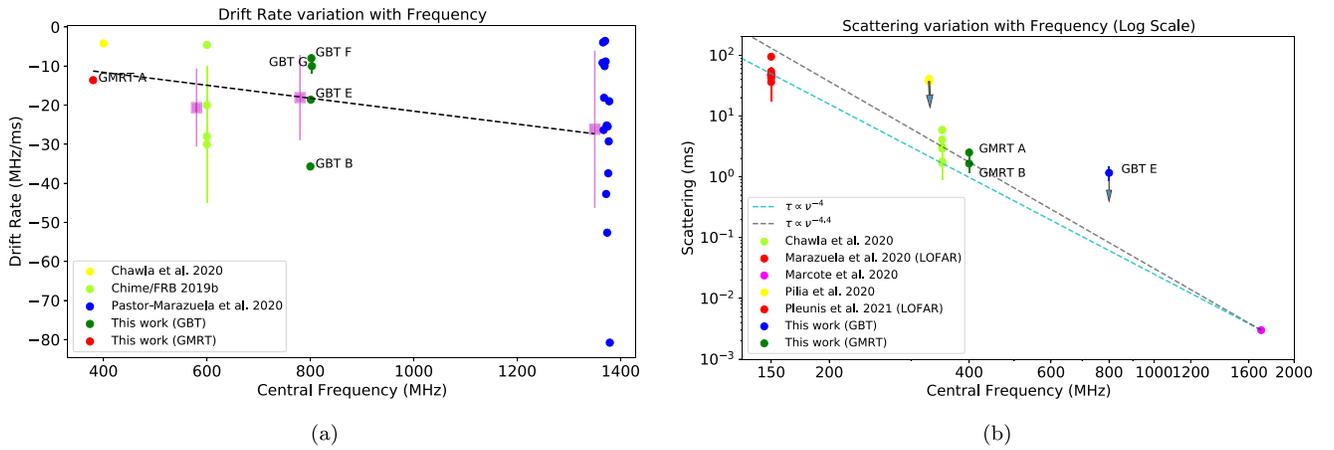

**Figure 7.** Comparison of drift rates and scattering as a function of observing frequencies for FRB 20180916B. (a) Measured and reported drift rates at different central frequencies. A least-squares fit across the mean value of the drift rate at each frequency (magenta boxes with error bars showing the standard deviation) gives us a slope of −0.02, i.e., $\dot{\nu} = -0.02\nu$, indicating that the drift rate appears to increase with observing frequency. (b) Measured and reported scattering timescale as a function of observing frequencies (shown in a log scale). Assuming a power law of −4 or even −4.4, GMRT scattering timescales agree with expected scattering extrapolated from the measured scattering timescale by Marcote et al. (2020) at a higher frequency. Data points with an arrow signify upper limits on scattering timescales since unresolved structures in the burst might lead to overestimation of scattering.

such structures in our bursts in both the time and frequency domain. Figure 7(b) shows scattering estimates for FRB 20180916B at different observing frequencies. It seems to follow a power law with an index of −4 extending from the 2.7 μs estimate by Marcote et al. (2020) using the scintillation bandwidth. This bandwidth estimate is in agreement with what is expected from the NE2001 model (Cordes & Lazio 2002). This hints that a majority of the scattering shown by the source across all frequencies can be attributed to the Milky Way, suggesting minimal contribution from the host. Scattering measurements for the bursts in our data set (Table 3) also concur with this trend. However, unresolved structures in the bursts can inhibit an accurate estimation of scattering timescales, leading to deviation from the expected value, as seen for GBT E in Figure 7(b). Here we show the upper limit on the scattering timescale of the burst (GBT E) since its precise estimation was hindered by its complex morphology. To understand these microstructures we performed a time-series autocorrelation analysis on high-S/N GBT bursts (GBT B, GBT E, and GBT F; see Section 3.1). We found GBT F showing structure around 30 μs wide (Figure 4(a)). This corresponds to an emission region of ∼10 km (using light travel time), hinting at a magnetospheric origin. In FRB 20180916B, structures have been seen down to 2 μs width by Nimmo et al. (2020) at 1700 MHz. The observation of such structures at even lower frequencies (<1 GHz) here suggest a cleaner local environment, not resulting in broadening and scattering of these tiny variations in the burst envelope.

Downward-drifting subbursts are a characteristic feature of repeaters. The slope of this "sad trombone" effect seems to increase with observing frequency. Hessels et al. (2019) first identified this phenomenon in FRB 20121102A bursts by comparing the drift-rate estimates from bursts detected at 1.4 GHz and 6 GHz. This trend has also been observed for FRB 20180916B by Pastor-Marazuela et al. (2021). As shown in Figure 7(a), our drift-rate estimates for a GMRT burst (GMRT A) and GBT bursts (GBT B and GBT E), measured in the same activity cycle of the source, do follow a linear trend with observing frequency. In Figure 7(a), we applied a best-fit slope, taking into consideration the mean of the drift rates (magenta boxes) for all bursts at each observing frequency. Though the large scatter in the measured drift rate (see Figure 7(a)) does suggest varying rates from burst-to-burst fluctuations at similar frequencies. To better understand the extent of this variation we estimated frequency-normalized drift rates i.e., drift rate/$\nu((ms)^{-1})$. On comparing this normalized drift rate with observed frequency we find the distribution of normalized drift rates to have a similar spread across all observing frequencies, although more data points are needed, especially in lower frequencies.

Overall, this correlation study for other repeating FRBs will be fundamental in modeling the emission mechanism shown by repeaters. Recent studies have also revealed a relation between the drift rates and temporal widths of individual subbursts for repeating sources (Chamma et al. 2021). This was not possible to investigate for our observations as the S/N of the detected subbursts and their temporal separation were not sufficient enough (except GMRT A; see Figure 3). However, we were able to resolve spectral striations and able to measure scintillation bandwidth (Section 3.4), which is in agreement with what is expected from the Milky Way (Cordes & Lazio 2002).

### 4.2. Phase–Frequency Activity Relation

Frequency dependence in the activity of FRB 20180916B is now well established (Aggarwal et al. 2020; Pastor-Marazuela et al. 2021; Pleunis et al. 2021). As shown in Figure 8, higher frequency detections occur earlier within the activity phase, with LOFAR detections occurring much later in the activity phase. We also see detections beyond the predicted "activity" window within the CHIME/FRB (CHIME/FRB Collaboration et al. 2020b), shown as the gray region (0.37–0.64 in activity phase). Pastor-Marazuela et al. (2021) found that the window seems to get narrower with increase in observing frequency and the burst rate varies from cycle to cycle, as was seen in their observations with the Apertif telescope. Estimation of burst rate with frequency at different phases over multiple cycles will help quantitatively estimate the activity of the source in both spectral and temporal phases. Overall, on comparing the peak activity we do see a downward trend in frequency with phase, i.e., lower frequencies appear to have later peak activity.

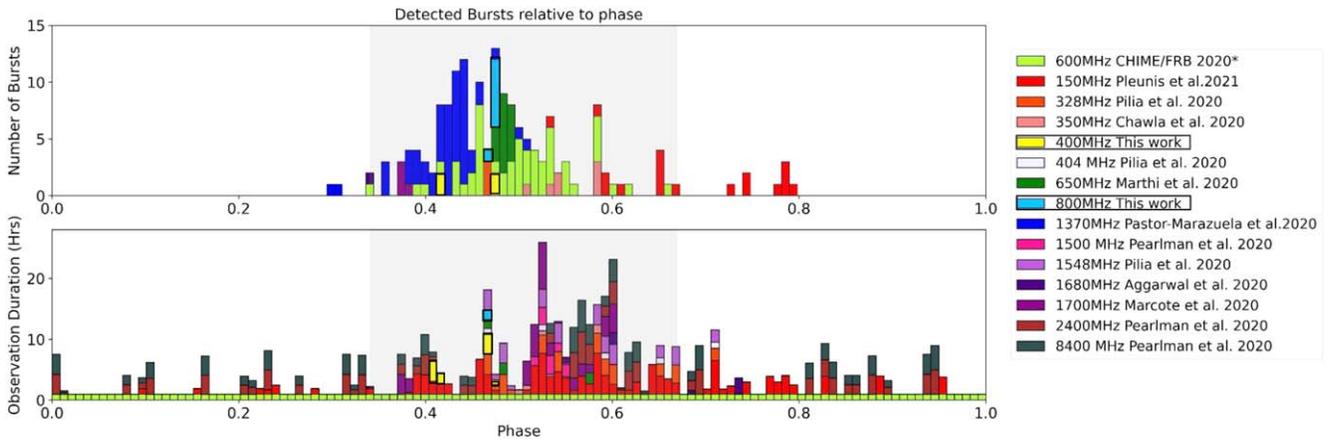

**Figure 8.** Frequency-dependent activity window of FRB 20180916B. Top: phase distribution of detected bursts with respect to frequencies folded at a period of 16.33 days with reference to MJD 58369.40 (Pleunis et al. [2021](#)). We see a clear shift in peak activity as we move from higher to lower observing frequencies. Bottom: observation hours at different phases for all observed frequencies. Here we see searches have been done beyond the activity window to constrain the periodicity, notably higher frequencies yielding no detections. We report here CHIME/FRB detections up to 2020 September 20. The gray shaded region shows the activity window. Observational data for APERTIF are not publicly available and hence are not shown in the bottom plot.

Our simultaneous GBT and GMRT observation epoch (0.46–0.48 in activity phase) perfectly follows this trend. We detected seven bursts from the source in 1.67 hr of observation with the GBT (800 MHz); this puts the burst rate to be $4.2^{+4.4}_{-2.5}$ bursts per hour above the fluence limit of 0.2 Jy ms with a 95% confidence limit. This rate is similar to what was observed by Marthi et al. ([2020](#)) at 550–750 MHz around a similar epoch. Interestingly, while the source was active in the GBT band, there were no detections in the uGMRT band (400 MHz), but we did detect two bursts in the same observing session (see Figure [1](#)). Hence, with a total of 2.85 hr of on-source time we get a burst rate of around $0.7^{+1.8}_{-0.6}$ bursts per hour above the 0.5 Jy ms fluence limit in the simultaneous observing window.

The fluence limits for both telescopes correspond to 2 ms wide bursts (narrowest detections in our sample) with $S/N > 6$ (single-pulse search limit) detection assuming a burst spectrum equal or smaller than our observed bandwidths. The GBT rate is 6 times that of GMRT, which might be due to our observations being in the peak activity of the source at 800 MHz. This can be seen in Figure [8](#) as GBT bursts seem to lie between the "peak" activity of Apertif (1370 MHz) and CHIME/FRB (600 MHz) detections. Such a high rate can also be serendipitous as clustering of bursts in arrival times have already been seen for FRB 20121102A (Gajjar et al. [2018](#); Zhang et al. [2018](#); Li et al. [2021](#)). The more interesting aspect is the complete lack of detections at lower frequencies (400 MHz) in the simultaneous time window, which might hint that the FRB remains active in one frequency range at a time. This can be due to beam geometry alignment with our line of sight or something intrinsic to the emission mechanism itself.

This one instance is not enough to conclusively comment on this frequency-dependent clustering. Multiband observations such as this one with burst-rate estimates spanning different phases will provide a more definitive picture; this is a topic for future research.

### 4.2.1. Frequency-dependent Spectral Variation

Spectrally limited emissions have now been well associated with repeaters (Kumar et al. [2020](#); CHIME/FRB Collaboration et al. [2021](#)). We too observe narrow emissions with varying central frequency in GMRT, GBT, and CHIME bands, although GMRT A and GMRT B cover almost the entire bandwidth. There is a lack of simultaneous emission as is evident from Figure [2](#). This restricts an exact estimation of the spectral index. To understand the energy distribution and its frequency dependence, Houben et al. ([2019](#)) characterized the statistical spectral index $\alpha_s$, which takes into consideration the statistical variability of the spectrum (or energy distribution) at different frequencies depending upon the frequency-dependent burst rate. The relation derived is as follows:

$$\frac{\lambda_1}{\lambda_2} = \left(\frac{\nu_1}{\nu_2}\right)^{-\alpha_s \gamma} \left(\frac{F_{\nu_1,\min}}{F_{\nu_2,\min}}\right)^{\gamma+1}. \qquad (2)$$

Here, $\gamma$ is the power-law index of the fluence distribution, which is taken to be $-2.3^{+0.4}_{-0.4}$ as estimated by the CHIME/FRB Collaboration et al. ([2020b](#)). $\lambda_1$ and $\lambda_2$ correspond to burst rates at central frequencies $\nu_1 = 400$ MHz and $\nu_2 = 800$ MHz, respectively. $F_{\nu,\min}$ is taken to be 0.9 times the fluence of the weakest burst detected (Houben et al. [2019](#)). In this case it is 1.34 Jy ms for GMRT ($0.9 \times$ GMRT D) and 0.22 Jy ms for GBT ($0.9 \times$ GBT D). After performing 10,000 simulations by randomly sampling values of $\lambda_1$, $\lambda_2$, and $\gamma$, assuming a Gaussian distribution using the mean and error range given above, we get $\alpha_s = -0.6^{+1.8}_{-0.9}$ with a reported 95% confidence limit.

Our measured statistical spectral index for FRB 20180916B is slightly higher than what was already reported by Chawla et al. ([2020](#)), although it agrees well within error bars. This slight discrepancy can be explained by an unusually high burst rate at the GBT band in just 1.67 hr of observation. This can be serendipitous or point to something intrinsic about the source. Notably, our observations are also at an earlier active phase than what was reported by Chawla et al. ([2020](#)), which can be suggestive of a phase-dependent energy distribution similar to the phase-dependent active emission frequency as mentioned in the previous section. This varying emission at different central frequencies still needs to be understood. Possible explanations involve something intrinsic to the emission region or a characteristic of propagation due to various phenomena such as plasma lensing or a refractive local medium (Cordes et al. [2017](#)). It is to be noted that due to our moderate burst sample and observing time our rates and thresholds provide limiting estimates to $\alpha_s$. Hence, more such simultaneous observations and estimation of the statistical spectral index across the activity window will aid better constraints of the distribution of burst energetics and possible emission mechanisms.

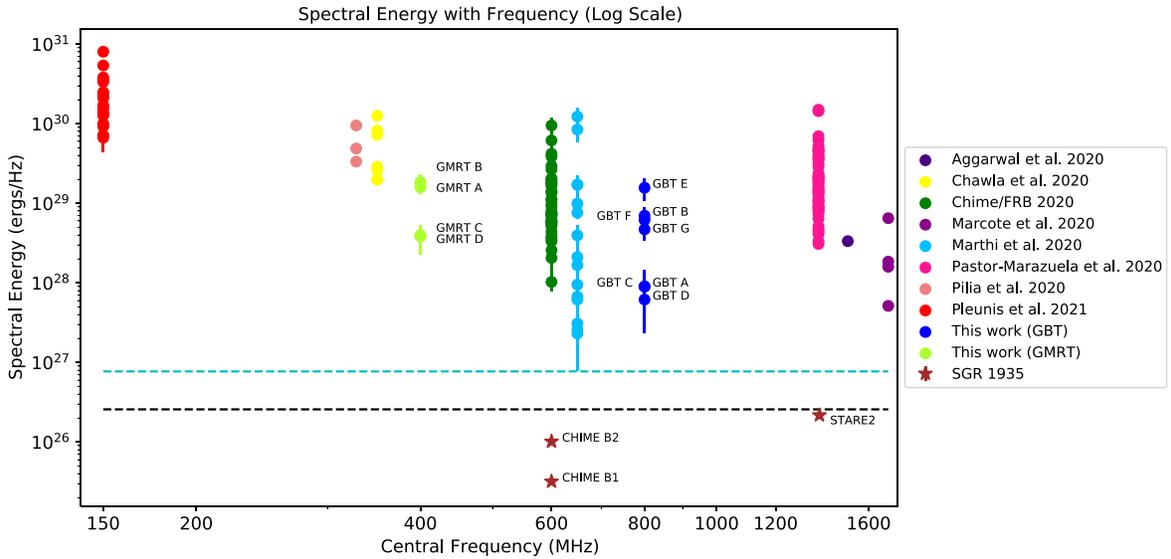

**Figure 9.** The spectral energy density of bursts as a function of observing frequencies for FRB 20180916B. We see bursts at lower frequencies are brighter due to higher sky temperature permitting only the strongest detections, although it is interesting to see a lack of energetic detections at >300 MHz. We see the weakest detections from FRB 20180916B are within an order of magnitude to the spectral energy of SGR J1935+2154 (shown by dotted lines) although the maximum possible distance to the source has been taken to estimate the energy (11 kpc). The distance to FRB 20190916B is taken to be 149 Mpc (Marcote et al. 2020). The blue dashed line shows the lower limit of the weakest burst from FRB 20180916B and the black dashed line shows the upper limit for the brightest detection from SGR 1935+2154. We clearly see that they differ by an order of magnitude, assuming we have correct distances.

### 4.3. Burst Energetics

FRB 20180916B is one of the closest repeating FRB sources to have been localized with an accurate distance estimate. Its frequent activity makes it ideal to study the energy budget of the source across multiple wavelengths in the radio regime. With our statistical spectral index estimation in Section 4.2.1 we found that there might be a phase–frequency dependence in energy distribution for the source. By comparing the energetics of individual bursts we can cumulatively estimate the energy released by the source at certain activity phases at different frequencies. Multiband observations such as this are ideal for such studies. This can help better understand the emission mechanism at play and narrow down the possible progenitor scenarios responsible for these events.

Figure 9 shows the spectral energy distribution of bursts detected from FRB 20180916B at various central frequencies and also makes a comparison with the energies of bursts from SGR 1935+2154. The spectral energy spans four orders of magnitude ranging from $10^{27}$ to $10^{31}$ ergs Hz$^{-1}$. The source appears to produce brighter bursts at lower frequencies, which, however, can be attributed to selection bias due to higher sky temperatures at these frequencies. Yet, nondetection of such highly energetic emissions at >300 MHz does put some constraint on the emission scenarios. Pastor-Marazuela et al. (2021) also mention that the FRB seems to be more active in the LOFAR band compared to the Apertif band at a fluence threshold of 50 Jy ms. We also detect brighter bursts at the GMRT band (400 MHz) compared to the GBT band (800 MHz). This is independent of GMRT's sensitivity since our fluence limit for the telescope (Section 4.2) is comparable to fluences for GBT bursts (Table 2), but with only four bursts in our data set we cannot make any conclusive claims. It is noteworthy, though, that Pearlman et al.'s (2020) emission limits at the *S* band (2.3 GHz–0.26 Jy ms) and *X* band (8.4 GHz–0.14 Jy ms) are comparable to the weakest detections observed in our case. Hence, the lack of high-frequency emission (>2 GHz) is likely to be intrinsic to the source, meaning the source can be active but not energetic enough at a frequency range, which can lead to no or very few detections at that frequency range, potentially affecting the periodicity estimates.

This brings the spectral energy budget in the radio wavelength to within an order of magnitude of what has been observed for the FRB detected from SGR 1935+2154. It is noteworthy that this burst was detected across a wide band, which includes 1.4 GHz detection by STARE2 (Bochenek et al. 2020) and detection down to 400 MHz by CHIME/FRB (CHIME/FRB Collaboration et al. 2020a), with emission possibly extending beyond these observing bands. This is equivalent to a bandwidth of more than 1000 MHz and, by taking the spectral energy estimate to be ∼$10^{26}$ ergs Hz$^{-1}$, we get total energy to be around $10^{35}$ ergs, which is comparable to the total energy of a burst from FRB 20180916B. This is due to the fact that in some cases the bandwidths of bursts from FRB 20180916B are <50 MHz (as for GBT A in our case), which amounts to energy in similar orders of magnitude. However, we should note that this is only true if we assume the narrow bandwidth emission is intrinsic to the source. The other caveat here is a large uncertainty in the distance to SGR 1935 + 2154 (Kothes et al. 2018; Zhong et al. 2020a; Zhou et al. 2020b), which can affect the energy argument by a few orders of magnitude. We can also see that SGR bursts (FRB 20200428) will not be detected by any of our telescopes if they are at the distance of FRB 20180916B; this does put into perspective the number of weaker detections we are missing at different frequencies, which might also affect the periodicity estimate. Narrowband emissions or flux knots have now been observed in giant pulses of PSR J0540-6919 (Geyer et al. 2021) and the Crab pulsar (Thulasiram & Lin 2021), which have been attributed to the intrinsic emission behavior of the source. Moreover, the discovery of a FRB in a globular cluster near M81 (Bhardwaj et al. 2021; Kirsten et al. 2022; Majid et al. 2021) does put forward a strong case for the relationship between giant pulses and FRBs (Nimmo et al. 2022). Another outburst comparable to FRB luminosity from a Galactic source will surely help us better constrain the energy gap between these two phenomena.

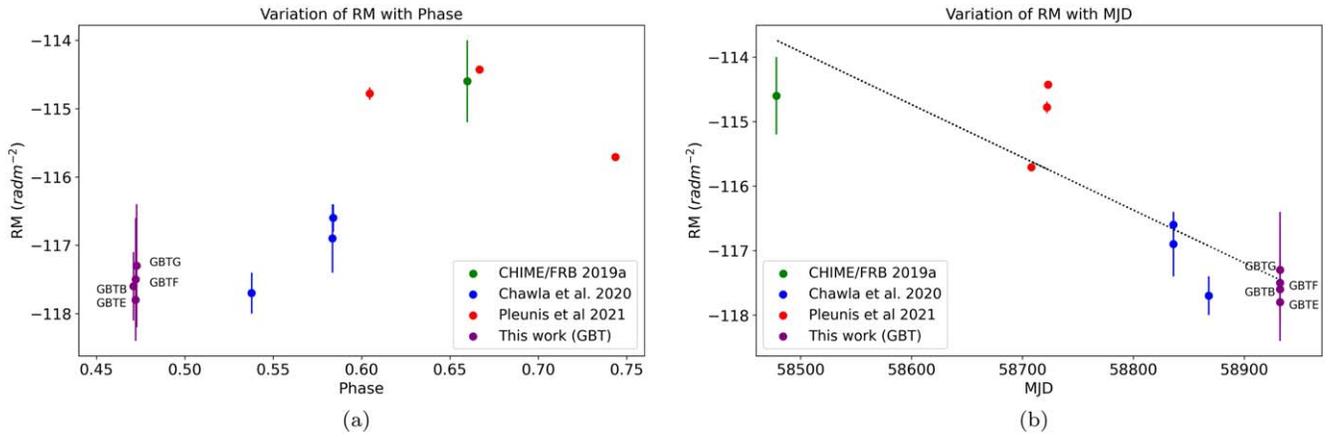

**Figure 10.** RM comparison for FRB 20180916B across multiple epochs and activity phases. (a) Folding the RM values as a function of activity phase for all observed epochs. Data is insufficient to observe any correlation in RM with phase; more observations in future might help uncover any underlying trend. (b) Observed RM as a function of observing epoch. The RM of the source gradually decreases since the first measurement by the CHIME/FRB Collaboration et al. (2019b) nearly two years ago. Our RM values (around $-117.5$ rad m$^{-2}$) are a unit less than the one reported by Chawla et al. (2020) for their 2019 November epoch ($-116.6$ rad m$^{-2}$) detections but similar to the one reported for their 2020 January detections ($-117.7$ rad m$^{-2}$).

### 4.4. Dispersion Measure and Rotation Measure Variations

DM variability is a strong indicator of changes in the local environment of the source, providing meaningful constraints on emission mechanisms as well as possible progenitor scenarios. FRB 20121102A showed a DM increase of $\sim$1–3 pc cm$^{-3}$ (Hessels et al. 2019; Li et al. 2021), supporting the hypothesis of the source being embedded in a dense nebula (Michilli et al. 2018). No such variation has been observed in the case of FRB 20180916B since its detection by CHIME/FRB. The initial estimated value of 348.82 pc cm$^{-3}$ (CHIME/FRB Collaboration et al. 2020b) has been reproduced throughout in subsequent detections. Nimmo et al. (2020) provided a strongest constraint of $348.772 \pm 0.006$ pc cm$^{-3}$ using micro-second time resolution. Our brightest detection, GBT E, has a structure-maximizing DM of $348.9 \pm 0.1$ pc cm$^{-3}$, which is in agreement with the reported value. On inspection by eye, the dynamic spectra of all the strong drifting bursts in our data set, namely GMRT A, GBT B, GBT E, and GBT F, showed maximum structures around 348.8 pc cm$^{-3}$. Other bursts seem to be weak or too fuzzy to accurately estimate their structure-maximizing DMs. Thus, we can constrain DM variation to be <1 pc cm$^{-3}$. We also did not observe any DM evolution between two GMRT epochs. This disfavors binary wind models and also supports a cleaner environment in the case of FRB 20180916B (Pastor-Marazuela et al. 2021).

We report the polarization properties of bursts detected from observations with the GBT. As shown in Figure 3, all bursts are 100% linearly polarized with no circular polarization. The PPA is approximately flat for each burst, and following calibration we do not observe any considerable variability in PPA offsets (see Table 3). This constant flat PPA is similar to what has been observed for FRB 20121102A (Hessels et al. 2019); however, the RM values between the two sources differ by four orders of magnitude (Michilli et al. 2018; CHIME/FRB Collaboration et al. 2019b). Also unlike FRB 20121102A, which shows a prominent decreasing trend in its RM with time (Hilmarsson et al. 2021), the RM changes slowly for FRB 20180916B. As seen in Figure 10, the RM has decreased over the two years ($\delta \sim$ 2–3 rad m$^{-2}$). This change can be due to alterations in the plasma environment or electron density in the local environment but, seeing the homogeneity in the DM, variations in viewing geometry can better explain this difference. We cannot comment on short-timescale variability (over a month) in RM values due to lack of data, but it would certainly be an interesting avenue to explore with a dedicated observing campaign. It is possible that, similar to burst activity, RM might also have a phase relation; however, we do not see a clear phase dependence (see Figure 10). More observations around different phases within the same activity window can uncover any likely dependence. Moreover, any variations in RM over short timescales (and phase) has to be removed to probe the true nature of RM variation over longer timescales. This is difficult to do with the limited sample but may be pursued in the future.

### 4.5. Progenitor Models

Numerous models have been put forward to explain the unique periodicity seen in FRB 20180916B, the most prominent being a pulsar/magnetar with a binary companion (Popov 2020; Zhang & Gao 2020), which could be an O/B star (Ioka & Zhang 2020; Lyutikov et al. 2020). In such a scenario, wind from the companion obscures the FRB source for most of the orbit. This model also favors frequency-dependent activity, with wider activity windows at higher frequencies. We see more activity in the GBT band (800 MHz) in comparison to the GMRT band (400 MHz) during our semisimultaneous observations. This early phase activity at higher frequencies has also been observed by Pastor-Marazuela et al. (2021), although they have shown a narrower activity window at higher frequencies, which disfavors such binary wind models. In addition to this, we also did not observe any DM evolution in our observations during the two GMRT observing sessions, challenging another prediction made by wind models.

Models supporting a precessing neutron star or magnetar have also tried to explain this periodicity (Levin et al. 2020; Tong et al. 2020). This framework requires a young magnetar surrounded with dense plasma; however, recent LOFAR detections (Pastor-Marazuela et al. 2021; Pleunis et al. 2021) and our energetic detections at the GMRT band (400 MHz) along with observation of microstructures (Figure 4) at the GBT band (800 MHz) favor a cleaner environment. Tendulkar et al. (2021) have also challenged many claims by these models with a high-resolution study of a 60 pc region around the localized position of FRB 20180916B using the Hubble Space Telescope. They instead favor a model involving a high-mass X-ray binary (HMXB) or a gamma-ray binary with a late O-type or B-type companion. Another interesting scenario that has tried to explain this periodicity of the source and in fact its frequency dependence involves emission of FRBs in ultra-luminous X-ray binaries (ULXBs; Deng et al. 2021; Sridhar et al. 2021) via a synchroton maser mechanism. But this model fails to explain the microsecond structures observed at different observing frequencies, as seen by Nimmo et al. (2020) at 1700 MHz and in our GBT observations at 800 MHz (see Figure 4). In summary, our low-frequency detections and constant PPA (see Section 3.3) are less likely to support a precessing young magnetar scenario inside a dense environment.

## 5. Conclusion

We detected 12 bursts (four, one, and seven) from FRB 20180916B in our 700 MHz wide observation using the uGMRT (300–500 MHz), CHIME (400–800 MHz), and GBT (600–1000 MHz), spread across 0.1 phase in the activity window. This includes simultaneous observation of the source by the GBT and uGMRT. Our reported first detections of the source in the 800–1000 MHz range indicates prolific activity of the source across 110–1700 MHz. Since our observations were earlier than the estimated peak activity (i.e., phase 0.5) we find the source to be more active in the GBT band compared to the GMRT band, which further reinforces earlier indications of a frequency-dependent activity window. Our estimates on the statistical spectral index also hint at the source being more energetic at higher frequencies earlier in its activity cycle. The observation of microstructure below 1 GHz here also provides evidence of a magnetospheric origin and a clean environment that does not lead to scattering and dissipation of these high-time-resolution variations in the burst envelope. The scattering seems to agree with what we can expect from our Milky Way although unresolved substructures do induce uncertainties in our estimates. The drift rate seems to increase with frequency similar to what has been observed for FRB 20121102A. We also report first polarimetric observations at 800 MHz, which is consistent with previous studies as we observe a 100% linear polarization and a flat PPA. We do find that the source RM has decreased over time, although not as significantly as FRB 20121102A, which is likely due to a four orders of magnitude difference in the initial RM between these two sources. We do not find any significant DM variation and constrain $\Delta$ DM $\sim 1$ pc cm$^{-3}$. With our results we disfavor the binary wind model as well as a precessing magnetar, but we have a limited phase coverage to make any significant assertions. Our results show that multiband studies of repeaters, or even one-off FRBs, is the ideal approach to disentangle any frequency dependence in the properties and activity of the source. Dedicated multiwavelength studies of FRB 20180916B across the entire activity window could be an exciting avenue for future research.

We thank the anonymous referee for their comments that have allowed us to improve our manuscript. We thank Avinash Deshpande, Wael Farah, Akshay Suresh, and Gaurav Waratkar for their helpful insights. We thank the staff of the GMRT that made the uGMRT observations possible. The GMRT is run by the National Centre for Radio Astrophysics of the Tata Institute of Fundamental Research.

We acknowledge that CHIME is located on the traditional, ancestral, and unceded territory of the Syilx/Okanagan people. We thank the Dominion Radio Astrophysical Observatory, operated by the National Research Council Canada, for gracious hospitality and expertise. CHIME is funded by a grant from the Canada Foundation for Innovation (CFI) 2012 Leading Edge Fund (Project 31170) and by contributions from the provinces of British Columbia, Québec, and Ontario. The CHIME/FRB Project is funded by a grant from the CFI 2015 Innovation Fund (Project 33213) and by contributions from the provinces of British Columbia and Québec, and by the Dunlap Institute for Astronomy and Astrophysics at the University of Toronto. Additional support was provided by the Canadian Institute for Advanced Research (CIFAR), McGill University and the McGill Space Institute via the Trottier Family Foundation, and the University of British Columbia.

Breakthrough Listen is managed by the Breakthrough Initiatives, sponsored by the Breakthrough Prize Foundation (breakthroughinitiatives.org). The Green Bank Observatory is a facility of the National Science Foundation operated under cooperative agreement by Associated Universities, Inc.

J.T.F. was generously supported by the National Science Foundation under the Berkeley SETI Research Center REU Site grant No. 1950897. A.B.P. is a McGill Space Institute (MSI) Fellow and a Fonds de Recherche du Quebec—Nature et Technologies (FRQNT) postdoctoral fellow. E.P. acknowledges funding from an NWO Veni Fellowship.

*Facilities:* GMRT, GBT, CHIME.

*Software:* astropy (Astropy Collaboration et al. 2013), DM phase (Seymour et al. 2019), DSPSR (van Straten & Bailes 2011), Gptool (Susobhanan et al. 2020), Heimdall (Barsdell et al. 2012), ionFR (Sotomayor-Beltran et al. 2013), Matplotlib (Hunter 2007), Numpy (Harris et al. 2020), PRESTO (Ransom 2011), PSRCHIVE (Hotan et al. 2004; van Straten et al. 2011), Scipy (Virtanen et al. 2020), SIGPROC (Lorimer 2011), SPANDAK (Gajjar et al. 2018).

## ORCID iDs

Ketan R. Sand  https://orcid.org/0000-0003-3154-3676
Jakob T. Faber  https://orcid.org/0000-0001-9855-5781
Vishal Gajjar  https://orcid.org/0000-0002-8604-106X
Daniele Michilli  https://orcid.org/0000-0002-2551-7554
Bridget C Andersen  https://orcid.org/0000-0001-5908-3152
Bhal Chandra Joshi  https://orcid.org/0000-0002-0863-7781
Sanjay Kudale  https://orcid.org/0000-0002-6631-1077
Maura Pilia  https://orcid.org/0000-0001-7397-8061
Bryan Brzycki  https://orcid.org/0000-0002-7461-107X
Tomas Cassanelli  https://orcid.org/0000-0003-2047-5276

Steve Croft ● https://orcid.org/0000-0003-4823-129X
Biprateep Dey ● https://orcid.org/0000-0002-5665-7912
Hoang John ● https://orcid.org/0000-0001-5591-5927
Calvin Leung ● https://orcid.org/0000-0002-4209-7408
Ryan Mckinven ● https://orcid.org/0000-0001-7348-6900
Cherry Ng ● https://orcid.org/0000-0002-3616-5160
Aaron B. Pearlman ● https://orcid.org/0000-0002-8912-0732
Emily Petroff ● https://orcid.org/0000-0002-9822-8008
Danny C. Price ● https://orcid.org/0000-0003-2783-1608
Andrew Siemion ● https://orcid.org/0000-0003-2828-7720
Kendrick Smith ● https://orcid.org/0000-0002-2088-3125
Shriharsh P. Tendulkar ● https://orcid.org/0000-0003-2548-2926